\documentclass[preprint,aps]{revtex4}

\usepackage{graphicx}
\usepackage{amssymb,amsmath}
\usepackage{caption}

\begin{document}

\title{Spectroscopic evidence for negative electronic compressibility in a quasi-three-dimensional spin-orbit correlated metal}
%
%
%
\author{Junfeng He$^{1}$\footnote[1]{These authors contributed equally to this work.} , T. Hogan$^{1*}$, Thomas R. Mion$^{1}$, H. Hafiz$^{2}$, Y. He$^{3}$, J. D. Denlinger$^{4}$, S.-K. Mo$^{4}$, C. Dhital$^{1}$, X. Chen$^{1}$, Qisen Lin$^{1}$, Y. Zhang$^{5}$, M. Hashimoto$^{3}$, H. Pan$^{1}$, D. H. Lu$^{3}$, M. Arita$^{6}$, K. Shimada$^{6}$, R. S. Markiewicz$^{2}$, Z. Wang$^{1}$, K. Kempa$^{1}$, M. J. Naughton$^{1}$, A. Bansil$^{2}$, S. D. Wilson$^{1,7}$, Rui-Hua He$^{1}$}

\affiliation{
\\$^{1}$Department of Physics, Boston College, Chestnut Hill, MA 02467, USA
\\$^{2}$Physics Department, Northeastern University, Boston, MA 02115, USA
\\$^{3}$Stanford Synchrotron Radiation Lightsource $\&$ Stanford Institute for Materials and Energy Sciences, SLAC National Accelerator Laboratory, Menlo Park, CA 94025, USA
\\$^{4}$Advanced Light Source, Lawrence Berkeley National Laboratory, Berkeley, CA 94720, USA
\\$^{5}$International Center for Quantum Materials, Peking University, Beijing 100871, China
\\$^{6}$Hiroshima Synchrotron Radiation Center, Hiroshima University, Hiroshima 739-0046, Japan
\\$^{7}$Materials Department, University of California Santa Barbara, Santa Barbara, CA 93106, USA
}

\begin{abstract}

Negative compressibility is a sign of thermodynamic instability of open \cite{Negative2, Negative7, Negative8} or non-equilibrium \cite{excited1,excited2} systems. In quantum materials consisting of multiple mutually coupled subsystems, the compressibility of one subsystem can be negative if it is countered by positive compressibility of the others. Manifestations of this effect have so far been limited to low-dimensional dilute electron systems\cite{Kravchenko, Eisenstein, Li, Yu, ascience, Ilani}. Here we present evidence from angle-resolved photoemission spectroscopy (ARPES) for negative electronic compressibility (NEC) in the quasi-three-dimensional (3D) spin-orbit correlated metal (Sr$_{1-x}$La$_x$)$_3$Ir$_2$O$_7$. Increased electron filling accompanies an anomalous decrease of the chemical potential, as indicated by the overall movement of the deep valence bands. Such anomaly, suggestive of NEC, is shown to be primarily driven by the lowering in energy of the conduction band as the correlated bandgap reduces. Our finding points to a distinct pathway towards an uncharted territory of NEC featuring bulk correlated metals with unique potential for applications in low-power nanoelectronics and novel metamaterials.

\end{abstract}


\maketitle

Electronic compressibility is defined as $\varkappa_e=\frac{1}{n^2}\frac{\partial n}{\partial\mu}$, where $n$ is the carrier density and $\mu$ the chemical potential. It is a fundamental thermodynamic quantity  that contains important information about the single-particle band structure as well as many-body effects in a material. Such information is experimentally accessible through ARPES, which provides independent measures of $n$ and changes in $\mu$, as we shall demonstrate in this study.

Sr$_3$Ir$_2$O$_7$ is an insulator driven by the cooperative interplay between correlation effects and strong spin-orbit coupling of the Ir \textit{5d} electrons \cite{KimPRL, Moon, SOMott6, Slater1}. It crystallizes in a near-tetragonal structure containing two square IrO$_2$ planes per unit cell, with the planar projection of the 3D Brillouin zone (BZ) as indicated by the white dashed line in Fig. 1a \cite{Cao327_2, Dessau327, King327, VidyaNM}. Sr$_3$Ir$_2$O$_7$ exhibits a significantly stronger three-dimensionality compared to the cuprates in that the c-axis and in-plane resistivities in Sr$_3$Ir$_2$O$_7$ differ only by a factor of $\sim$8, whereas cuprates show a corresponding difference of two to four orders of magnitude \cite{Cao327_2}. Substituting one Sr atom with La effectively dopes one electron into the system, leading to an insulator-metal transition at $x> 0.03$ in (Sr$_{1-x}$La$_x$)$_3$Ir$_2$O$_7$ \cite{Cao327_2}. Our ARPES experiments (see Methods) show that the doped electrons initially populate the conduction band around its bottoms near the $M$ points, forming four apparent electron pockets at the Fermi level in each planar BZ (Fig. 1a). The number, momentum location, overall shape of the observed pockets and the associated band dispersions, all agree well with our first-principles calculations (compare Fig. 1a,c with d,g and Supplementary Fig. 1; see Methods). As $x$ increases, electron pockets expand in momentum, as indicated by an increasing separation between the two given Fermi momenta on the pocket (Fig. 1c,e). Values of both $n$ and the effective La substitution level $x_{ARPES}$ can be estimated for each measured sample on the basis of the known lattice constants \cite{chetanPRB} and the total percentage area of the Fermi surface pockets in the planar BZ, with a small uncertainty arising from the relatively weak band dispersion along the c-axis (see Supplementary Discussion 1 and Supplementary Figs 2 and 3). The $x_{ARPES}$ value determined via ARPES is consistent with the La substitution level measured using energy-dispersive x-ray spectroscopy (EDS) on the same sample (Fig. 1f).

Photoemission can measure relative changes in $\mu$ as a function of doping in a given material with reference to electronic states located sufficiently far away from the Fermi level such that their energies are largely unaffected by changes in doping (see Supplementary Discussion 2). Two types of electronic reference are generally used, the core levels and the (deep) valence bands located at a binding energy of $\sim 6$ eV, for which consistent results have been demonstrated experimentally \cite{kylePRL, YagiPRB}. Use of core levels as references requires an understanding of energy shifts due to various factors other than changes in $\mu$ itself \cite{Fujimori_2}. These other effects, however, are expected to be averaged out for deep valence bands, which adopt an increased itinerancy relative to the core levels (as exemplified by the substantial dispersion in Fig. 2a), and the associated energy shifts can then simply represent changes in $\mu$ (Supplementary Discussion 2). Our observations in (Sr$_{1-x}$La$_x$)$_3$Ir$_2$O$_7$ are consistent with this deep-valence-band approach, where the general shape of the spectra containing multiple orbital components remains qualitatively unchanged with doping (Fig. 2a,b,d,e). For example, as $x$ increases, we see that, in Fig. 2d, the O \textit{2p} orbital at the $X$ point \cite{King327} systematically decreases in binding energy, revealing in turn a decrease of $\mu$. A similar behavior was found for different states, such as those around $\Gamma$, which are of a mixed Ir-O character and exhibit an unambiguous overall upward band shift (Fig. 2a,b,e). Relative shifts in $\mu$ deduced from various deep valence band states observable under our experimental conditions are summarized in Fig. 2c, and consistently show a systematic decrease of $\mu$ with increasing electron filling. This trend is also in qualitative accord with limited results obtained by using core-level references (Supplementary Fig. 4 and Supplementary Discussion 2).

In a rigid-band picture, all bands exhibit the same shift in binding energy with fixed inter-band separations. Electron doping in such a case would necessarily lead to an increase of $\mu$ and, hence, a positive electronic compressibility. Rather, we observed a decrease of $\mu$, indicative of a NEC, in metallic (Sr$_{1-x}$La$_x$)$_3$Ir$_2$O$_7$. This suggests a breakdown of the rigid-band picture and necessitates changes in the separations of certain bands in the system (and their associated `internal' energy). To obtain insights into the physical mechanism underlying the observed NEC, we thus turn our attention away from the high-binding-energy region in which bands appear `rigid' to search for outliers accountable for the NEC. As we shall show, these are the conduction and (shallow) valence bands located at low binding energies, which exhibit non-trivial movements with doping.

With increasing electron filling, the conduction band bottom near $M$ sinks deeper under the Fermi level (Fig. 3c), whereas the valence band top at $X$ rises (Fig. 3a,b), resulting in a decrease of the associated (indirect) bandgap (Fig. 3d). Closer inspection suggests that the conduction and valence band movements are not only momentum dependent (Supplementary Fig. 1) but also asymmetric, as shown in Fig. 4a, where we have combined the information concerning the doping dependence of $\mu$ (Fig. 2c) with that of the conduction and shallow valence bands. The conduction band bottom is seen to decrease continuously (by $\sim$100 meV) as $x$ increases from $0.035$ to $0.086$, whereas the apparent valence band top exhibits an abrupt increase in energy (by $\sim$150 meV) across $x\sim 0.05$, and an otherwise moderate (if any) shift away from this doping region. This abrupt variation indicates a change in character of the dominant states defining the apparent valence band top. The observed spectral evolution across $x\sim 0.05$ is consistent with a spectral weight transfer from the original valence band states to new states which appear at lower energy inside the original bandgap (Supplementary Fig. 5g). This type of spectral weight transfer is typical of electron-doped transition metal oxides \cite{CorrelatedOxideReview2}, including Sr$_{2-x}$La$_x$IrO$_4$ \cite{Opt214_Tokura}.

The marked overall reduction of bandgap, appearance of in-gap states, and spectral weight transfer over a wide energy range, all with moderate electron doping, point to the correlated nature of the low-lying states close to $\mu$. The overall situation is similar to that of electron-doped cuprates, Nd$_{2-x}$Ce$_x$CuO$_4$ \cite{Armitagereview} (see Fig. 4b,c), where the bandgap between the lower-lying (in-gap) shallow valence band and the conduction band is understood as being due to antiferromagnetic correlations, and the higher-lying (original) shallow valence band is a remnant of the lower Hubbard band \cite{Armitagereview, Bansil_1}. Whether or not the bandgap in metallic (Sr$_{1-x}$La$_x$)$_3$Ir$_2$O$_7$ is magnetic in nature, we note the possibility that the $J=1/2$ antiferromagnetic correlations at $x=0$, driven by on-site Coulomb repulsion and spin-orbit coupling \cite{KimPRL, Moon, SOMott6, Slater1}, might persist over a short range in the metallic regime and weaken with increasing doping, as in Nd$_{2-x}$Ce$_x$CuO$_4$ \cite{Armitagereview}.

However, a reduction of the correlation gap is not a sufficient basis for understanding the counterintuitive shift in $\mu$ observed in (Sr$_{1-x}$La$_x$)$_3$Ir$_2$O$_7$, as $\mu$ was found to evolve in the normal way with doping in Nd$_{2-x}$Ce$_x$CuO$_4$ \cite{Fujimori_2, Armitagereview} (see Fig. 4b,c). Another essential aspect is the manner in which the bandgap decreases. In general, the Fermi level tends to move up in energy relative to the conduction band bottom, reflecting increased electron filling, whereas it is insensitive to the movements of the shallow valence band as long as it stays occupied. But, in (Sr$_{1-x}$La$_x$)$_3$Ir$_2$O$_7$, the conduction band bottom itself moves down rapidly as the bandgap shrinks, resulting in an effective lowering of $\mu$. In contrast, in Nd$_{2-x}$Ce$_x$CuO$_4$, the conduction band bottom barely moves with $x$ instead, the valence band top moves upwards rapidly.

Therefore, the unique evolution of the correlation gap holds the key to understanding our observed NEC. Several microscopic aspects make the case of (Sr$_{1-x}$La$_x$)$_3$Ir$_2$O$_7$ distinct from Nd$_{2-x}$Ce$_x$CuO$_4$. First, as a result of a negative $\frac{\partial n}{\partial\mu}$, local electron-density fluctuations will promote electronic phase separation, which would be frustrated by the long-range Coulomb interactions and confined to a microscopic scale \cite{SOC_LAOSTO}. Although in-gap states can exist in systems without microscopic phase separation (for example, in Nd$_{2-x}$Ce$_x$CuO$_4$ \cite{Armitagereview}), their presence along with a nanoscale phase separation in La$_{2-x}$Sr$_x$CuO$_4$ might be involved in the doping independence of $\mu$ observed in a certain doping range \cite{Fujimori_2, NegCompCuprate}. Whether microscopic phase separation exists in metallic (Sr$_{1-x}$La$_x$)$_3$Ir$_2$O$_7$ warrants further study. Second, strong spin-orbit coupling in the iridates is responsible for the formation of an effective $J=1/2$ band from a mixture of three $t_{2g}$ orbitals, enabling a putative Mott transition to take place in this band without the need for strong on-site Coulomb repulsion \cite{KimPRL, Moon, SOMott6, Slater1}. It has been pointed out that a large inter-orbital charge transfer with NEC is feasible in a multi-band model with at least one band being close to a Mott transition \cite{Mannhart}. On the other hand, weakening electron correlations may lead to an effective spin-orbit coupling strength that decreases with electron doping \cite{SOC_SRO}. Interestingly, an electron-density dependence of one type of spin-orbit coupling was proposed to be essential for the NEC found at an oxide heterointerface \cite{SOC_LAOSTO}. Last, a unique form of structural distortion in (Sr$_{1-x}$La$_x$)$_3$Ir$_2$O$_7$ contributes to the bandgap formation (Supplementary Fig. 6), and may bring about other subtle changes in the electronic structure via deformation potentials \cite{DeformationPotential}. The extent to which the strong spin-orbit and/or multi-orbital nature of the correlated electronic states, as well as the structural effects, are at play in their non-trivial doping-dependent movements should be examined to understand the microscopic driving force underlying the NEC in (Sr$_{1-x}$La$_x$)$_3$Ir$_2$O$_7$.

Our finding suggests that the quasi-3D electron system in metallic (Sr$_{1-x}$La$_x$)$_3$Ir$_2$O$_7$ represents the first experimental case of NEC in three dimensions, which has been discussed theoretically \cite{Mannhart, CompSumRule2} as an important complement to the lower-dimensional cases that have been found in the conventional \cite{Kravchenko, Eisenstein} or oxide \cite{Li} semiconductor heterojunctions, monolayer \cite{Yu} and bilayer \cite{ascience} graphene, and carbon nanotubes \cite{Ilani}. The earlier lower-dimensional cases were established mainly by probing $\varkappa_e$ via quantum capacitance \cite{Kravchenko, Li, Yu, Ilani} or electric field penetration \cite{Eisenstein, Li}, both quantities being proportional to $\frac{\partial n}{\partial\mu}$. The application of these techniques for investigating (Sr$_{1-x}$La$_x$)$_3$Ir$_2$O$_7$ single crystals will be complicated by the relatively high value of $|\frac{\partial n}{\partial\mu}|$
($=5.7 \times 10^{18}$ meV$^{-1}$cm$^{-3}$) deduced from ARPES, the requirements of surface flatness and/or thickness in the nanoscale for related device fabrications, and the need to tune carrier concentration during measurements; it would probably be more promising to work with thin-film samples, which have yet to be grown. In any case, our ARPES study adds 3D NEC to the growing list of unusual properties of the iridates.

We have presented evidence of a new pathway for realizing NEC in which the electron correlation energy dominates, whereas in all other known cases of NEC, it is the exchange energy that dominates \cite{Kravchenko, Eisenstein, Li, Yu, ascience, Ilani}. Such a route could in principle apply to many correlated materials \cite{Mannhart}. Correlated metals such as doped transition metal oxides or dichalcogenides with strong spin-orbit coupling and/or the presence of multiple bands might provide a fertile ground for exploring novel NEC phenomena.

Bulk materials with NEC would lead to unique possibilities for research and applications. The compressibility sum rule \cite{CompSumRule2} implies that a NEC will lead to a negative dielectric constant near zero frequency. The bulk nature of NEC materials would thus allow the RF/optical study of $\varkappa_e$, and applications in metamaterials as active components, neither being feasible for their lower-dimensional counterparts. The use of metal electrodes with negative quantum capacitance in transistors could effectively enhance their gate capacitances, presenting an alternative to the use of `high-$\kappa$' dielectrics for miniaturization of devices that switch at low voltage with minimal gate-to-channel leakage \cite{Li}. In contrast to two-dimensional electron/hole systems \cite{Kravchenko, Eisenstein, Li}, doped correlated metals are typically associated with a high correlation energy scale (on the order of $ 100$ meV) \cite{CorrelatedOxideReview2}, and owing to their 3D bulk nature, they are amenable to changes of environment as well as deposition onto any substrate, creating devices with working resistance tunable by film thickness. Application of materials with NEC in transistors thus promises good adaptability to the existing CMOS architecture, and potentially enables room-temperature (field independent) and variable-frequency device operation.\\

\noindent{\textbf{Methods} }

\textbf{Experiment.} Single crystals of (Sr$_{1-x}$La$_x$)$_3$Ir$_2$O$_7$ with different $x$ were grown by flux techniques similar to earlier reports\cite{chetanPRB, Cao327_2}. Samples were cleaved at 30 K in ultrahigh vacuum before ARPES measurements. The presented ARPES results were obtained at 30 K, mostly at Beamline 5-4 of the Stanford Synchrotron Radiation Lightsource (SSRL) of SLAC National Accelerator Laboratory using 25 eV photons with a total energy resolution of $\sim$9 meV and a base pressure better than 3$\times$10$^{-11}$ Torr. Stability of the photon energy and energy position of the sample Fermi level was constantly monitored by measuring a polycrystalline Au reference sample in electrical contact with the iridate sample, which showed a typical variation $<\pm 1$ meV during the measurement on each sample. Fermi surface maps shown resulted from an integration over a $\pm 10$-meV window around the Fermi level. A related preliminary experiment was performed on the $x=0.08$ sample at Beamline 9 A of the Hiroshima Synchrotron Radiation Center (HSRC). Core-level and work function measurements were performed at Beamlines 10 \& 4 of the Advanced Light Source (ALS) at Lawrence Berkeley National Laboratory using 180 eV and 83 eV photons with total energy resolutions of $\sim$50 meV and $\sim$20 meV, respectively. A  related  preliminary  experiment was  performed  at  BL47XU of SPring-8 using 8 keV photons. As noted in Supplementary Discussions 2 and 3, variations of the sample work function do not affect the measured energy shifts of deep valence bands and core levels (Supplementary Fig. 7); limited results of work function measurements on (Sr$_{1-x}$La$_x$)$_3$Ir$_2$O$_7$ via photoemission are not inconsistent with a NEC (Supplementary Fig. 8).\\

\textbf{Calculation.} First-principles calculations were performed along the lines described in Ref. \cite{VidyaNM} using the Vienna Ab initio Simulation Package (VASP). The core and valence electrons were treated by the projector augmented wave (PAW) and a plane wave basis, respectively. Exchange-correlation effects were incorporated using the generalized gradient approximation (GGA). A $\sqrt{2} \times \sqrt{2}$ superlattice was used with an in-plane lattice constant of $3.9 \AA$ to take into account the $12^\circ$-rotation of the oxygen octahedra surrounding the Ir atoms. Antiferromagnetic order, when present, does not further change the size of the unit cell. Electron-electron interaction between the correlated $d$ electrons on Ir atoms was included at the GGA$+U$ (on-site Coulomb repulsion) mean-field level. To simulate the experimental results shown in Fig. 1, we set $U=1.1$ eV and increased the strength of spin-orbit coupling by a factor of two compared to the value obtained in self-consistent computations. The chemical potential is set in the calculations such that the Fermi surface volume of the conduction band is consistent with the electron filling.\\

\noindent{\textbf{Acknowledgements} } We thank K. S. Burch, T.-R. Chang, Y.-H. Chu, A. Fujimori, Z. Hussain, S. A. Kivelson, H. Lin, V. Madhavan, Z.-X. Shen, Q. Si, C. M. Varma, Z.-Y. Weng, H. Yao and X. J. Zhou for discussions, and K. Tanaka for measuring sample work functions at UVSOR, Japan. The work at Boston College was supported by a BC start-up fund (J.H., R.-H.H.), the US NSF CAREER Awards DMR-1454926 (R.-H.H., in part) and DMR-1056625 (T.H., C.D., X.C., S.D.W.), NSF Graduate Research Fellowship GRFP-5100141 (T.R.M.), DOE-DE-SC0002554 and DE-FG02-99ER45747 (Z.W.) and the W. M. Keck Foundation (M.J.N.). The work at Northeastern University (NU) was supported by the DOE, BES Contract No. DE-FG02-07ER46352, and benefited from NU's ASCC and the allocation of supercomputer time at NERSC through DOE grant DE-AC02-05CH11231. Photoemission experiments were performed at the SSRL and ALS, supported respectively by the US DOE BES Contract Nos. DE-AC02-76SF00515 and DE-AC02-05CH11231, at HSRC and SPring-8 (preliminary) with the approval of Proposal Nos. 14-A-1 and 2014B1501.\\

\noindent{\textbf{Competing Interests} } The authors declare that they have no competing financial interests.\\

\noindent{\textbf{Author Contributions} } J.H. and T.H. contributed equally to this work. J.H. and R.-H.H. proposed and designed the research. J.H. and T.R.M. carried out the ARPES measurements with help from Y.H., J.D.D., S.-K.M., Q.L. and H.P.. T.H., C.D. and X.C. grew the samples. T.H., J.H. and T.R.M. characterized the samples with EDS. H.H. and R.S.M. performed the first-principles calculations and, along with A.B. and K.K., provided theoretical guidance. M.H., D.H.L., S.-K.M., M.A. and K.S. maintained the experimental facilities. J.H. analyzed the data with help from Y.Z.. J.H. and R.-H.H. wrote the paper with key inputs from Z.W., Y.H., S.D.W., A.B., and R.S.M.. R.-H.H., S.D.W., M.J.N. and A.B. are responsible for project direction, planning and infrastructure.\\

\noindent{\textbf{Correspondence} }
Correspondence and requests for materials should be addressed to R.-H.H..\\


\newpage
\begin{figure*}[tbp]
\includegraphics[width=0.9\columnwidth,angle=0]{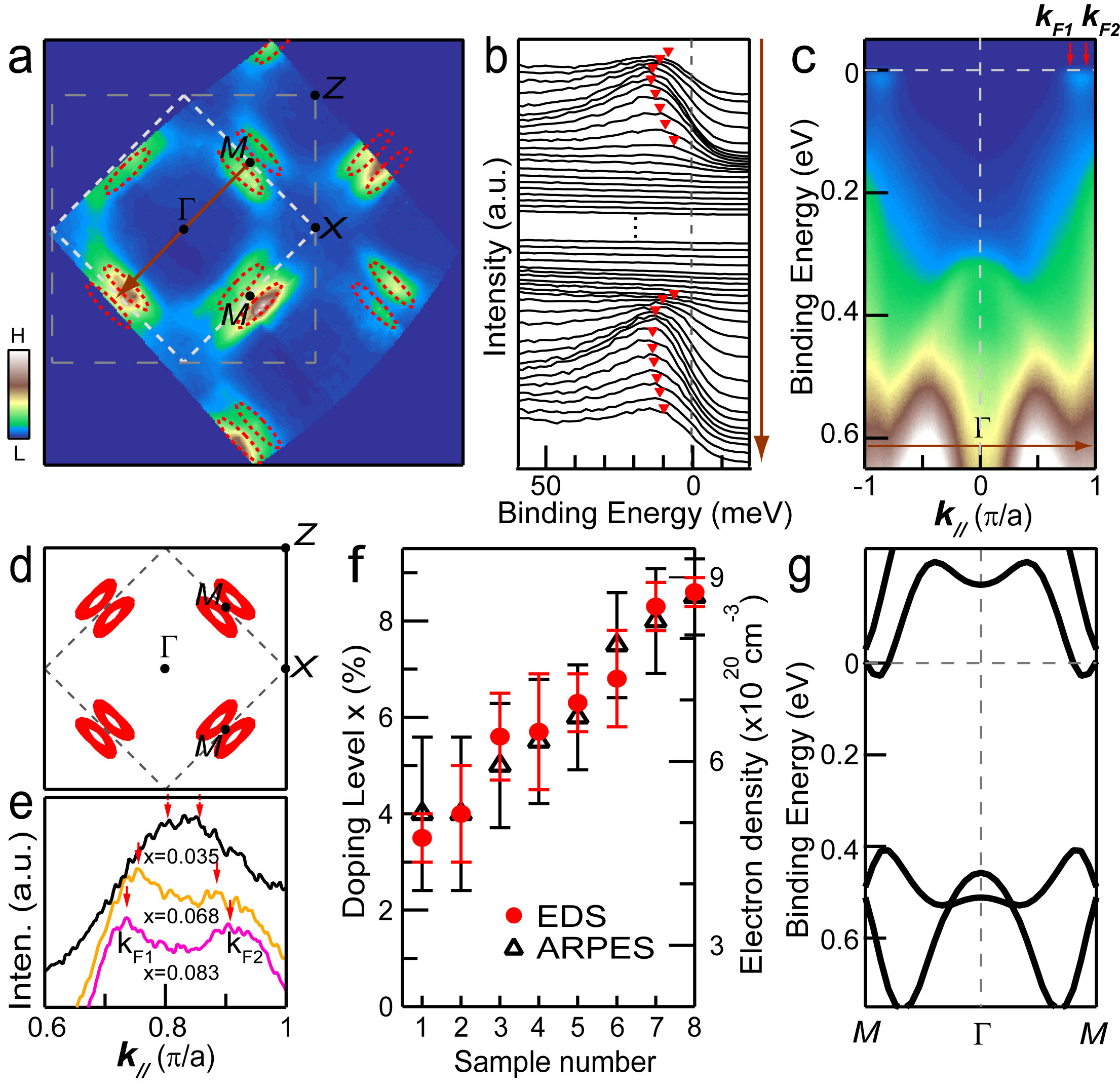}
\begin{flushleft}
\caption{\textbf{Fermi surface and band structure of (Sr$_{1-x}$La$_x$)$_3$Ir$_2$O$_7$ for $x=0.057$.} \textbf{a}, Fermi surface map featuring two electron pockets (red dashed ellipses) around each $M$ point. Brillouin zone boundary for the distorted (undistorted) lattice is marked by white (gray) dashed line. \textbf{b},\textbf{c}, Energy distribution curves (EDCs) (\textbf{b}) and band dispersion map (\textbf{c}) along the brown arrow in \textbf{a}, revealing two bands close to the Fermi level as traced by red triangles in \textbf{b}. $a=5.50 $ \AA. \textbf{d},\textbf{g}, Calculated Fermi surface (\textbf{d}) and band dispersion (\textbf{g}) in comparison with \textbf{a},\textbf{c}. \textbf{e}, Momentum distribution curves at the Fermi level for selected $x$ values, showing peaks at $k_{F1}$ and $k_{F2}$ as indicated by the red arrows (also shown in \textbf{c}). \textbf{f}, La substitution level and the corresponding electron density, determined by EDS (red circles) and ARPES (black triangles) for various samples studied. Error bars reflect the spatial variation on a given sample surface (EDS) and uncertainty in determining the Fermi momentum (ARPES).}
\end{flushleft}
\end{figure*}

\newpage

\begin{figure*}[tbp]
\includegraphics[width=1.0\columnwidth,angle=0]{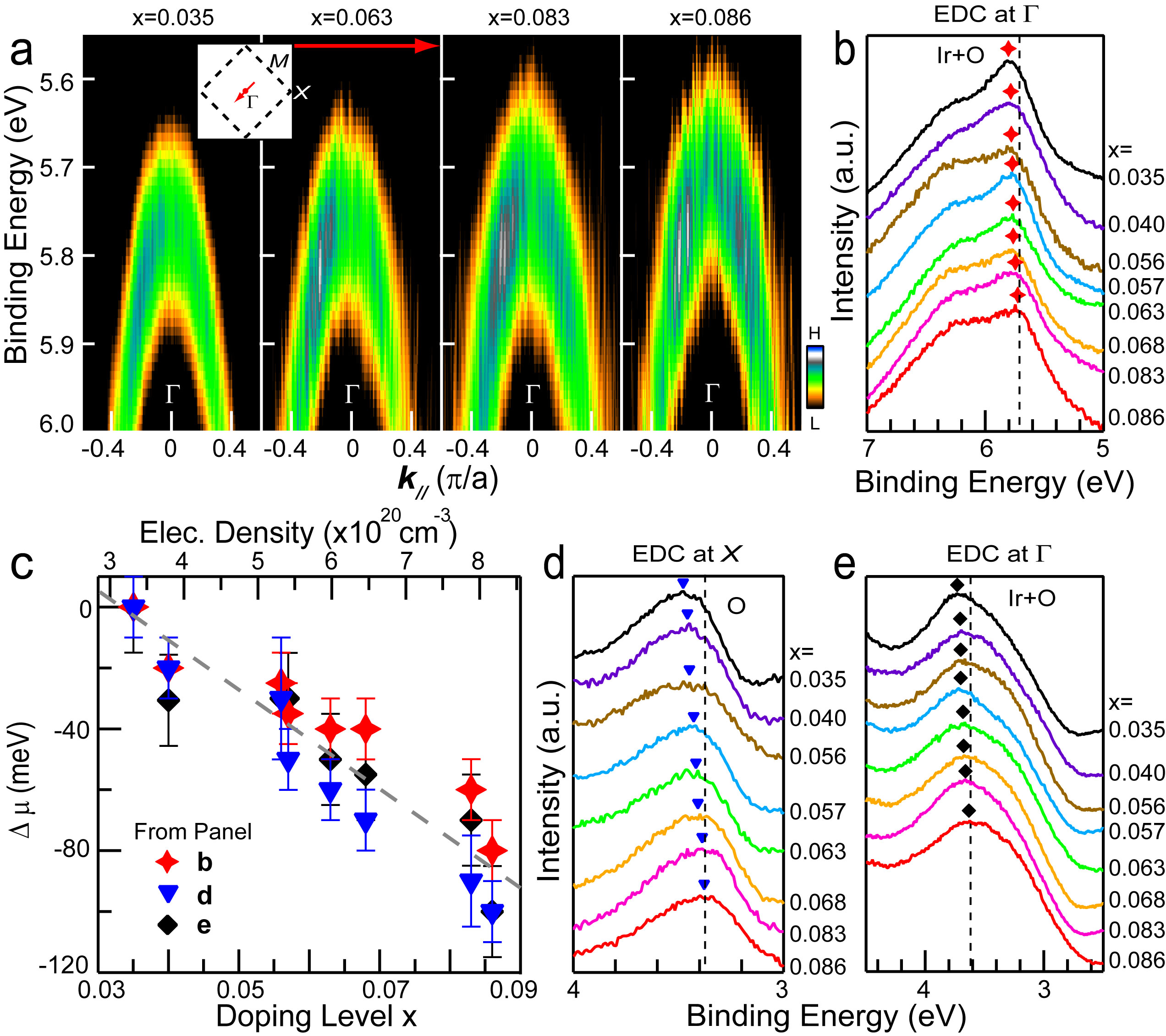}
\begin{flushleft}
\caption{\textbf{Chemical potential shift on electron doping in (Sr$_{1-x}$La$_x$)$_3$Ir$_2$O$_7$.} \textbf{a}, EDC-second-derivative band dispersion maps along the red arrow (inset) in the deep valence band region around the $\Gamma$ point for selected $x$ values. \textbf{b},\textbf{d},\textbf{e}, EDCs at $\Gamma$ (\textbf{b},\textbf{e}) and X (\textbf{d}) shown in different energy ranges for various $x$. Symbols indicate spectral peaks for states with a mixed Ir-O (in \textbf{b},\textbf{e}) or O 2p character (in \textbf{d}). \textbf{a} shows the dispersions of the peaks in \textbf{b}. See Supplementary Fig. 9 for assignments of orbital character to states based on calculations. \textbf{c}, Chemical potential shift (guided by the grey dashed line) as a function of $x$ and the calculated electron density, deduced from the corresponding states marked in \textbf{b, d, e}. The chemical potential for $x=0.035$ is set, for convenience, as a zero reference for the relative shift in chemical potentials. Error bars reflect the uncertainty in determining the EDC peak energy position.}
\end{flushleft}
\end{figure*}

\newpage
\begin{figure*}[tbp]
\includegraphics[width=1.0\columnwidth,angle=0]{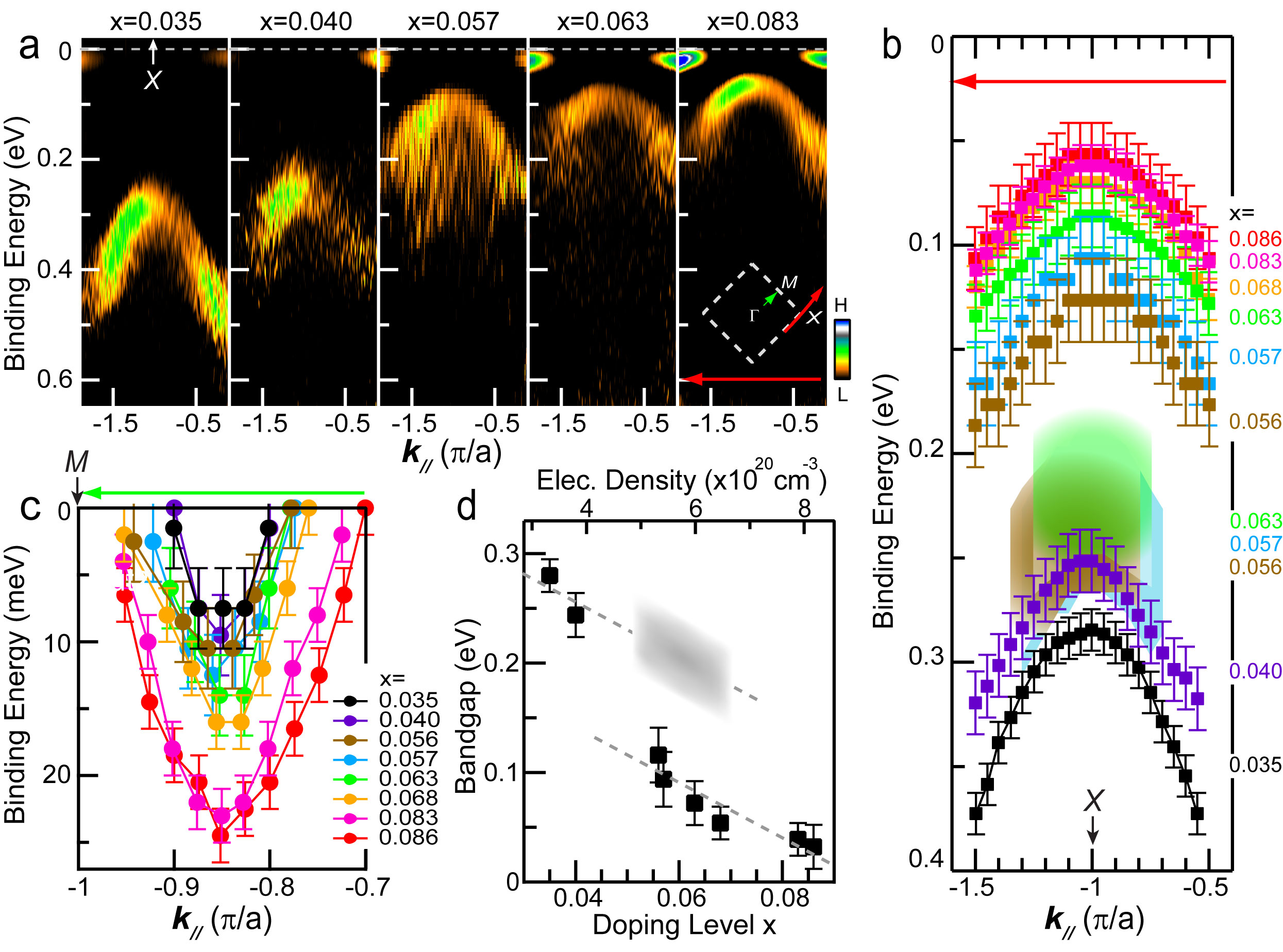}
\begin{flushleft}
\caption{\textbf{Doping evolution of the valence band top and conduction band bottom of (Sr$_{1-x}$La$_x$)$_3$Ir$_2$O$_7$.} \textbf{a}, EDC-second-derivative band dispersion maps of the shallow valence band along the red arrow (inset) for selected $x$ values. \textbf{b},\textbf{c}, Doping evolution of the extracted shallow valence band (\textbf{b}) and conduction band (\textbf{c}) dispersions along, respectively, the red and green arrows shown in the inset of \textbf{a}. Valence band spectra at $x=0.056$, $0.057$ and $0.063$ contain multiple features (see Supplementary Fig.5). \textbf{d}, Doping evolution of the indirect bandgap. The valence band top is defined by the dominant valence band feature in \textbf{a}, which changes character across $x\sim 0.057$. The apparent bandgap has accordingly different characters as guided by the dashed lines. Grey shaded area denotes the bandgap roughly defined with the higher-energy feature (color shaded area in \textbf{b}) of the shallow valence band at $x\sim 0.057$. Error bars reflect the uncertainty in determining the second-derivative EDC peak energy position.}
\end{flushleft}
\end{figure*}

\newpage
\begin{figure*}[tbp]
\includegraphics[width=1.0\columnwidth,angle=0]{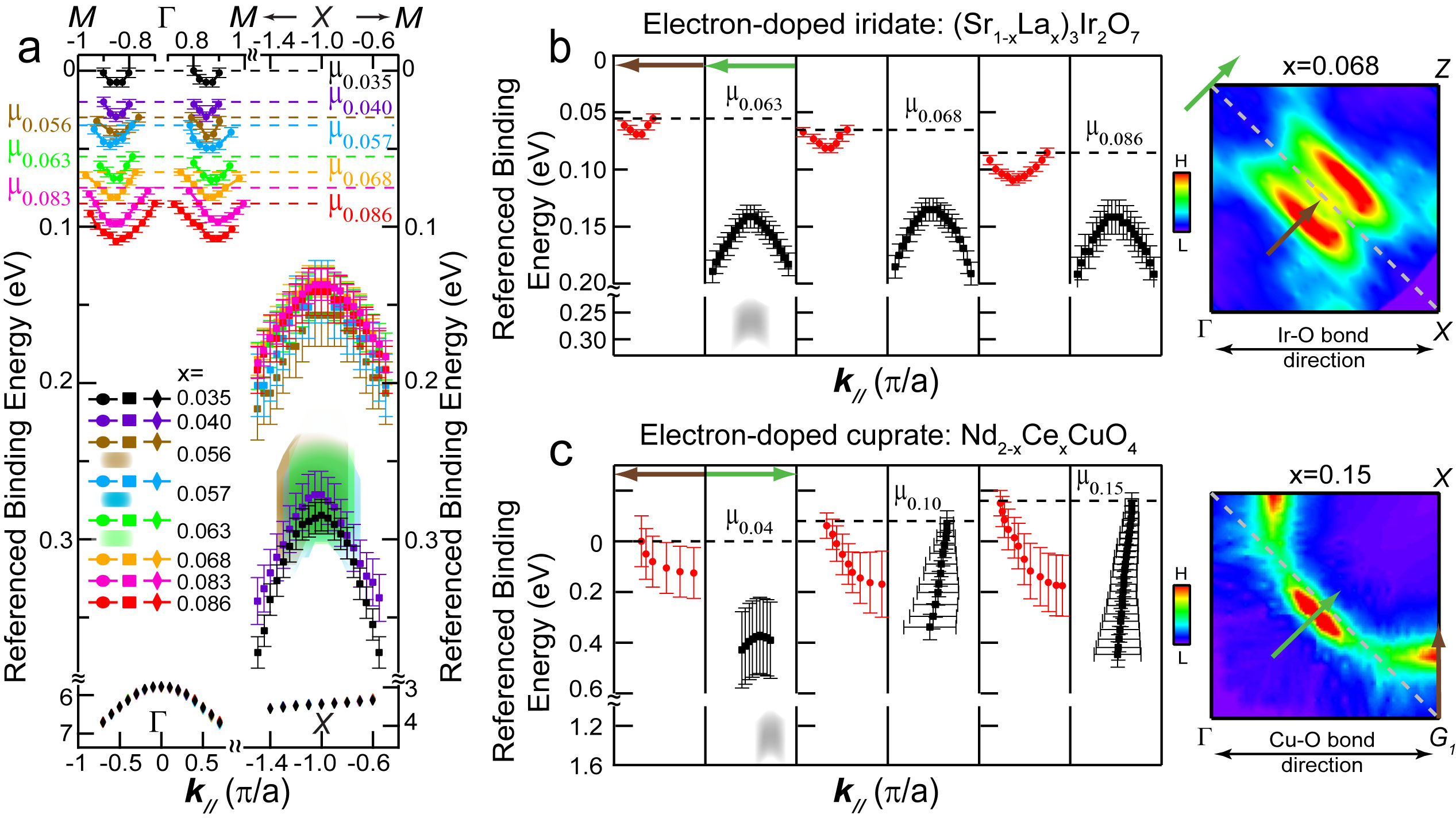}
\begin{flushleft}
\caption{\textbf{Doping evolution of the band structure of (Sr$_{1-x}$La$_x$)$_3$Ir$_2$O$_7$ and its comparison with Nd$_{2-x}$Ce$_x$CuO$_4$.} \textbf{a}, Extracted dispersions of the conduction band (circles; along the brown arrow in Fig. 1a), the shallow valence band (squares and shaded areas; along the red arrow in Fig. 3a), and selected deep valence bands (diamonds, largely overlapping; along both arrows), for various $x$ values. Results are plotted on a common energy scale relative to their respective chemical potentials (dashed lines), with zero defined by that of $x=0.035$ (Fig. 2c). \textbf{b},\textbf{c}, Doping evolution of the conduction band, shallow valence band and chemical potential for (Sr$_{1-x}$La$_x$)$_3$Ir$_2$O$_7$ and Nd$_{2-x}$Ce$_x$CuO$_4$ (after refs. \cite{Fujimori_2, Armitagereview}), respectively. Their conduction (red) and valence (black) bands are located at different momenta, as covered by the arrows in the respective (symmetrized) Fermi surface maps. The grey shaded areas denote possible higher-energy features of the shallow valence bands. Error bars are reproduced from those in Figs 2c and 3b,c and in ref.\cite{Armitagereview}.}
\end{flushleft}
\end{figure*}


%
%
%

\end{document}